\documentclass[showpacs,twocolumn,prl]{revtex4}
\usepackage{graphicx,psfrag,amssymb}
\newcommand{\ket}[1]{\left|#1\right\rangle}
\newcommand{\bra}[1]{\left\langle#1\right|}
\newcommand{\op}[2]{\ket{#1}\bra{#2}}
\newcommand{\proj}[1]{\op{#1}{#1}}
\newcommand{\com}[2]{\left[#1,#2\right]}
\newcommand{\brac}[1]{\left\{#1\right\}}

\newcommand{\brar}[1]{\left(#1\right)}
\newcommand{\bras}[1]{\left[#1\right]}
\newcommand{\dampingnobr}[3]{2#1#2#3-#3#1#2-#2#3#1} 
\newcommand{\damping}[3]{\brar{\dampingnobr{#1}{#2}{#3}}}
\newcommand{\modulus}[1]{\left|#1\right|}
\newcommand{\conj}[1]{{#1}^\ast}
\begin{document}
\title{Entanglement and Entropy Engineering of Atomic Two-Qubit
States}
\author{S.~G. Clark}
\author{A.~S. Parkins}
\email[Corresponding author. Email address: ]{s.parkins@auckland.ac.nz}
\affiliation{Department of Physics, University of Auckland,
Private Bag 92019, Auckland, New Zealand.}
\date{\today}

\begin{abstract}
We propose a scheme employing quantum-reservoir engineering to
controllably entangle the internal states of two atoms
trapped in a high finesse optical cavity.
Using laser and cavity fields to drive two separate Raman
transitions between metastable atomic ground states,
a system is realized corresponding to a pair of two-state atoms
coupled collectively to a squeezed reservoir.
Phase-sensitive reservoir correlations lead to entanglement
between the atoms, and, via
local unitary transformations and adjustment of the degree and
purity of squeezing,
one can prepare entangled mixed states
with any allowed combination of linear entropy and
entanglement of formation.
\end{abstract}
\maketitle

The properties of entangled mixed states and schemes for their
controlled preparation are presently under vigorous investigation,
primarily because of their relevance to understanding the role of
purity and entanglement in quantum information protocols such as
quantum computation and quantum communication \cite{Whi01}.
The purity and degree of entanglement of two-qubit states can be
quantified, respectively, by the linear entropy and either the
entanglement of formation or the concurrence \cite{Mun01}.
Here, we propose a scheme using interactions in cavity quantum
electrodynamics (cavity QED) which enables the preparation
of states of two atomic qubits with {\em any} allowed combination
of linear entropy and concurrence.

Our scheme uses the technique of quantum-reservoir engineering
\cite{Lut98} in a cavity QED setting to couple a pair of two-state
atoms collectively to an effective squeezed reservoir.
The phase-sensitive quantum correlations of the
reservoir are transferred to the two-atom system to produce
entangled atomic states \cite{Pal89,Aga89}.
The degrees of purity and entanglement
of the atomic states can be controlled through the excitation time,
through properties of the effective squeezing (i.e., the degree and
purity of squeezing), and through adjustment of the relative
strengths of amplitude and phase coupling to the reservoir.
We are thus able to scan the entire allowed region
of the linear entropy--concurrence plane (or
linear entropy--tangle plane \cite{Mun01}), including the region
between the Werner states \cite{Wer89} and the recently characterized
maximally-entangled mixed states (MEMS) \cite{Mun01}.
Other cavity-QED-based schemes for entangling a pair of atoms
have been proposed and even implemented
(see, e.g., \cite{Hag97,Zhe00,Ple99}), but these schemes have focussed
primarily on generating maximally-entangled pure states.

\begin{figure}\begin{psfrags}
  \psfrag{Dr}{$\Delta_r$}
  \psfrag{Ds}{$\Delta_s$}
  \psfrag{Dt}{$\Delta_t$}
  \psfrag{Or}{$\Omega_r$}
  \psfrag{Os}{$\Omega_s$}
  \psfrag{Ot}{$\Omega_t$}
  \psfrag{1}{$\ket{0}$}
  \psfrag{2}{$\ket{r}$}
  \psfrag{3}{$\ket{1}$}
  \psfrag{4}{$\ket{s}$}
  \psfrag{5}{$\ket{t}$}
  \psfrag{gg}{$g_r$}
  \psfrag{ge}{$g_s$}
  \psfrag{delta}{$\delta$}
\includegraphics[scale=0.4]{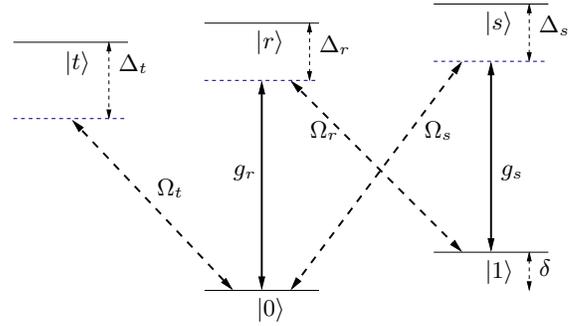}
\caption{Atomic level scheme for each atom. The excited states have
energies $\hbar\omega_j$ ($j=r,s,t$).
}
\label{fig:levels}
\end{psfrags}\end{figure}

In our proposal, two atoms are assumed to be tightly confined inside a
high-finesse optical cavity and separated by a distance that is
sufficiently large that they
can be individually addressed by probe lasers and so that there is
no direct dipole-dipole interaction between them.
The cavity has a field decay rate of $\kappa$ and a frequency
$\omega$, and may, if desired, be driven with broadband thermal
light characterized by a mean photon number $\bar{n}$.
Two stable ground states ($\ket{0},\ket{1}$)
of each atom constitute the qubit
states (Fig.~\ref{fig:levels}).
The cavity field and two auxiliary laser fields drive two
separate resonant Raman transitions between these states. In particular,
transitions $\ket{1}\leftrightarrow\ket{r}$ and
$\ket{0}\leftrightarrow\ket{s}$ are driven by detuned laser fields
with (real) Rabi frequencies $\Omega_r$ and $\Omega_s$ and relative
phase difference $\varphi$,
while the transitions $\ket{0}\leftrightarrow\ket{r}$
and $\ket{1}\leftrightarrow\ket{s}$ are strongly coupled to the
cavity mode, with coupling strengths $g_r$ and $g_s$
(assumed the same for both atoms).
Detunings of the fields from the excited states $\ket{r}$ and
$\ket{s}$ are given by $\Delta_r$ and $\Delta_s$.
A fifth state $\ket{t}$ is virtually excited from $\ket{0}$ by
another strongly detuned laser field, adding an additional
ac-Stark shift to the state $\ket{0}$.

The master equation for the total system density operator
is (taking $\hbar =1$)
\begin{equation}
  \dot{\rho}=-i\com{H}{\rho}+\mathcal{L}_{\rm cav}\rho
  +\mathcal{L}_{\rm spon}\rho ,
\end{equation}
where $H=H_{\rm cav}+H_{\rm at}+H_{\rm int}$, with
$H_{\rm cav}=\omega a^\dag a$,
\begin{eqnarray}
 H_{\rm at}&=&\sum_{i=1,2} \left\{ \,
  \omega_r\proj{r_i}+\omega_s\proj{s_i}+\omega_t\proj{t_i}
   \right.
\nonumber\\
  &+& \delta\proj{1_i} + [ (\Omega_r/2)
  \textrm{e}^{-i\omega_{L_r}t}\op{r_i}{1_i}+\textrm{H.c.} ]
\nonumber\\
  &+& [ (\Omega_s/2)
  \textrm{e}^{-i\bras{\omega_{L_s} t+\varphi}}\op{s_i}{0_i}+\textrm{H.c.} ]
\nonumber\\
  &+& [ (\Omega_t/2)
  \textrm{e}^{-i\omega_{L_t}t}\op{t_i}{0_i}+\textrm{H.c.} ] \left. \right\} ,
\nonumber\\
  H_{\rm int}&=&\sum_{i=1,2} \left( g_r\op{r_i}{0_i}a
  +g_s\op{s_i}{1_i}a+\textrm{H.c.} \right) ,
\end{eqnarray}
(H.c. denotes Hermitian conjugate) and
\begin{eqnarray}
  \mathcal{L}_{\rm cav}\rho&=&\kappa(1+\bar{n})
  \damping{a}{\rho}{a^\dag}
\nonumber\\
  &&+\kappa\bar{n}\damping{a^\dag}{\rho}{a} .
\end{eqnarray}
Here, $\omega_{Lj}$ ($j=r,s,t$) denote the laser frequencies.
The term $\mathcal{L}_{\rm spon}\rho$ describes atomic spontaneous
emission.

To isolate the essential dynamics, we
assume large detunings
of the light fields from the excited atomic states
(i.e.,
$\modulus{\Delta_j}\gg \Omega_j,g_r,g_s,\kappa,\gamma_j$,
where $\gamma_j$ is the linewidth of state $\ket{j}$),
so that atomic spontaneous emission is negligible and the excited
states can be adiabatically eliminated from the problem.
This leads to a reduced master
equation for a pair of effective two-level atoms (involving
states $\ket{0}$ and $\ket{1}$) coupled to the cavity mode.
This reduced system is characterized by the parameters
\begin{eqnarray}
  \beta_j=g_j\Omega_j/(2\Delta_j), \;\;
  \eta_j=g_j^2/\Delta_j, \;\; j=\{ r,s\},
\end{eqnarray}
where $\beta_r$ and $\beta_s$ are the two (Raman) coupling
strengths, and $\eta_r$ and
$\eta_s$ are the ac-Stark shifts per cavity photon induced in
$\ket{0}$ and $\ket{1}$, respectively.

To further reduce the model, we assume the ``bad-cavity''
limit,
$\kappa\gg\modulus{\beta_{r,s}},\modulus{\eta_{r,s}}$.
This enables us to adiabatically eliminate
the cavity mode, which yields a master equation
for the atomic density matrix in the form
\begin{eqnarray}
  \dot\rho&=&(2\beta^2/\kappa )(N+1)\damping{S}{\rho}{S^\dag}
\nonumber\\*
  &+&(2\beta^2/\kappa )N\damping{S^\dag}{\rho}{S}
\nonumber\\*
  &-&(2\beta^2/\kappa )M\damping{S^\dag}{\rho}{S^\dag}
\nonumber\\*
  &-&(2\beta^2/\kappa )\conj{M}\damping{S}{\rho}{S}
\nonumber\\*
  &+&(\eta^2/2\kappa )\bar{n}(\bar{n}+1)\damping{P}{\rho}{P^\dag}.
\label{eq:me}
\end{eqnarray}
Here, $\beta^2=\beta_r^2-\beta_s^2$, $\eta^2=\brar{\eta_r-\eta_s}^2$, and
\begin{eqnarray}
  N=\frac{\brar{\bar{n}+1}\beta_s^2+\bar{n}\beta_r^2}{\beta^2}, \;\;\;
  M=\frac{-\brar{2\bar{n}+1}\beta_r\beta_s\textrm{e}^{i\varphi}}{\beta^2} ,
\end{eqnarray}
while $S=\left(\sigma_1^-+\sigma_2^-\right)/\sqrt{2}$ and
$P=\sigma_1^-\sigma_1^++\sigma_2^-\sigma_2^+$ are collective atomic
operators, with
$\sigma^-_i=\op{0_i}{1_i}$.

The derivation of (\ref{eq:me}) also requires
that the phase of the effective two-level system remains constant
with respect to the laser phase difference $\varphi$.
That is, the effective atomic system and squeezed reservoir
must be ``resonant'' with eachother, which requires that
\begin{equation}
  \frac{\Omega_s^2}{4\Delta_s}-
  \frac{\Omega_r^2}{4\Delta_r}+
  \frac{\Omega_t^2}{4\Delta_t}+
  \frac{g_r^2}{\Delta_r}\bar{n}-
  \frac{g_s^2}{\Delta_s}\bar{n}=0.
  \label{eq:condition}
\end{equation}
It is to satisfy this condition
while retaining flexibility in our choices of $\Omega_{r,s}$ and
$\Delta_{r,s}$
that we use the additional transition
$\ket{0}\leftrightarrow\ket{t}$. The level shift
$\Omega_t^2/(4\Delta_t)$ provides an extra degree of
freedom with which to satisfy (\ref{eq:condition}).

In (\ref{eq:me}), the terms proportional to $\beta^2$
describe the collective (amplitude) coupling of
our pair of effective two-level atoms to an effective squeezed
reservoir, with the degree and purity of squeezing
characterized by the parameters $\{ N,M\}$ \cite{Pal89,Aga89}.
In particular, the effective squeezed quadrature variance is
proportional to $(N-|M|+1/2)$ and ideal squeezing corresponds to
$|M|^2=N(N+1)$, which requires that $\bar{n}=0$.
The last line of (\ref{eq:me}) describes
phase damping of the atomic qubits
caused by coherent scattering of off-resonant (thermal)
intracavity photons.
There is no phase damping if $\bar{n}=0$ or if $\eta_r=\eta_s$.

A feature of the present system is that the strengths
of the amplitude and phase damping terms are independently
adjustable, so that, for example, one can be made to dominate the
other
(remembering that (\ref{eq:condition}) must remain satisfied).
Also, by switching off all sources of light (i.e., setting
$\beta_r=\beta_s=0$ and $\bar{n}=0$) the state of the two-atom
system can in principle be ``frozen'' at any instant.

To begin our analysis of (\ref{eq:me}),
we note first that associated with
the collective coupling of the atoms to the reservoir
are certain decoherence-free states, which decouple completely
from the dynamics \cite{Ple99}. In particular, defining
$\ket{\phi^\pm}=\brar{\ket{00}\pm\ket{11}}/\sqrt{2}$ and
$\ket{\psi^\pm}=\brar{\ket{01}\pm\ket{10}}/\sqrt{2}$,
one finds that $\ket{\psi^-}$ decouples for all parameter
choices, while $\ket{\psi^+}$ decouples if $N=M=0$.

As a first example, we consider the case in which phase
damping can be neglected
(i.e., $\eta =0$ or $\eta^2\ll\beta^2$).
The steady state density matrix $\rho_{\rm ss}$ is then,
assuming an initial state that has no projection onto $\ket{\psi^-}$,
given by
\begin{equation} \label{eq:rhoss}
  \rho_{\rm ss}=\brar{\begin{array}{cccc}
  \rho_{11}&0&0&\rho_{14}\\
  0&\rho_{22}&\rho_{23}&0\\
  0&\rho_{32}&\rho_{33}&0
  \\\rho_{41}&0&0&\rho_{44}\\\end{array}},
\end{equation}
specified in the basis $\brac{\ket{11},\ket{10},\ket{01},\ket{00}}$,
with
\begin{eqnarray}
  \rho_{11}=\frac{|M|^2(1-2N)+
  {N}^2(1+2N)}{(1+2N)L} \ ,  \;\;\;\;\;
\\
  \rho_{22}=\rho_{33}=\rho_{23}=\frac16-\frac{1}{6L} , \;\;
  \rho_{14}=\frac{M}{(1+2N)L} \ ,
\end{eqnarray}
where $L=1+3N(1+N)-3|M|^2$.

\begin{figure}\begin{psfrags}
  \includegraphics[scale=0.4]{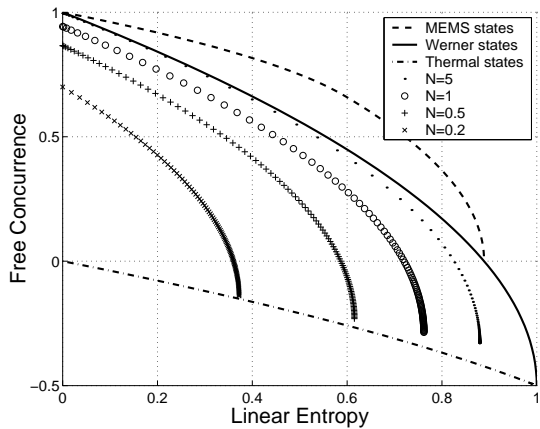}
  \caption{Steady state values of ${\cal C}_{\rm free}(\rho )$ and
  $S_{\rm L}(\rho )$ for selected values of $N$ and
  $0\le |M|^2\le N(N+1)$, for initial state $\ket{00}$.}
  \label{fig:steady}
\end{psfrags}\end{figure}

\begin{figure}\begin{psfrags}
  \includegraphics[scale=0.4]{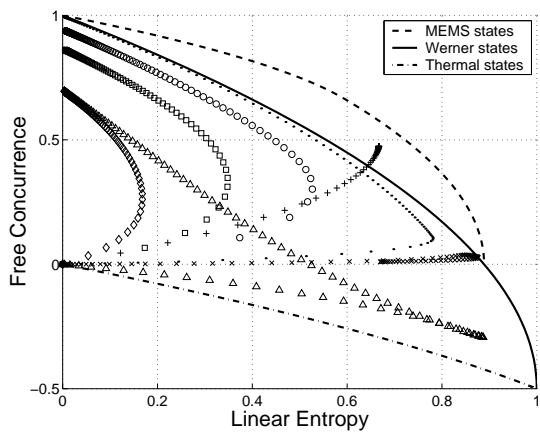}
  \caption{Evolution of ${\cal C}_{\rm free}(\rho )$ and
  $S_{\rm L}(\rho )$ to the steady state with ideal
  squeezing ($|M|^2=N(N+1)$) for the following
  initial states and values of $N$:
  $\{\ket{00},N=0.2\}$ ($\diamond$),
  $\{\ket{00},N=0.5\}$ ($\square$),
  $\{\ket{00},N=1\}$ ({\large $\circ$}),
  $\{\ket{00},N=5\}$ ($\bullet$),
  $\{\ket{11},N=0.2\}$ ($\triangle$),
  $\{\ket{01},N=2\}$ ($\times$),
  $\{\ket{01},N=0.01\}$ ($+$).
  Note that the points on each curve
  are not equally spaced in time.}
  \label{fig:driven}
\end{psfrags}\end{figure}

Examples of
the steady state and time evolution to the steady state are shown in
Figs.~\ref{fig:steady} and \ref{fig:driven}, respectively, plotted
as points in the linear entropy-concurrence plane
\cite{EoF}. We plot a
minor variation of the true definition of the concurrence and call it the
{\it free concurrence}
${\cal C}_{\rm free}=(\lambda_1-\lambda_2-\lambda_3-\lambda_4)$, where
$\lambda_{1-4}$ are the square roots of the eigenvalues, in decreasing
order, of $\rho\tilde{\rho}$, where
$\tilde{\rho}=(\sigma_{y}\otimes\sigma_{y})\rho^\ast
(\sigma_{y}\otimes\sigma_{y})$,
with $\sigma_{y}=-i(\sigma^+-\sigma^-)$.
The use of the free concurrence
enables separable states (${\cal C}_{\rm free}\le0$) to be more readily
distinguished. The linear entropy is given by
$S_{\rm L}(\rho )=(4/3)[1-{\rm Tr}(\rho^2)]$.
On each graph we also plot lines corresponding to the Werner states,
$\rho_W=\xi\proj{\phi^+}+(1/4)(1-\xi )\openone_4$
($0\leq\xi\leq 1$), the MEMS
of~\cite{Mun01}, which have the maximum amount of entanglement
for a given linear entropy, and thermal states
$\rho_{\rm th}=[\zeta\ket{0}\bra{0}+(1-\zeta )\ket{1}
\bra{1}]^{\otimes 2}$ ($0\leq\zeta\leq 1$).
For ideal squeezing the steady state described by
(\ref{eq:rhoss}) is the pure state \cite{Pal89}
\begin{equation}
  \ket{\Psi_{\rm s}}=\sqrt{\frac{N+1}{1+2N}}\ket{00}
  -{\rm e}^{i\varphi}\sqrt{\frac{N}{1+2N}}\ket{11}.
\end{equation}
These states lie on the left-hand vertical axes of the figures
and approach the Bell states $\ket{\phi^\pm}$
in the limit of large squeezing (i.e., large $N$, and $\varphi =\pi$
or $0$).
Nonideal squeezing (Fig.~\ref{fig:steady}) generates steady states that
can lie essentially anywhere below the Werner line.
Note that for large $N$ the steady states
closely approximate mixtures of $\proj{\phi^+}$ ($\varphi =\pi$) and
$\rho^\prime =\textrm{diag}\brac{1/3,1/6,1/6,1/3}$.

Time evolution with ideal squeezing from initial states with zero
projection onto $\ket{\psi^-}$ can also sweep out the region beneath
the Werner line (Fig.~\ref{fig:driven}).
When the initial state {\em does} have a projection onto $\ket{\psi^-}$
(e.g., $\ket{01}$),
an interesting range of points on the plane can also be accessed,
including an area above the Werner line and the region along the
boundary at ${\cal C}_{\rm free}=0$ between separable and entangled
(including the maximally mixed entangled state at the intersection of the
Werner and MEMS lines).

\begin{figure}\begin{psfrags}
  \includegraphics[scale=0.4]{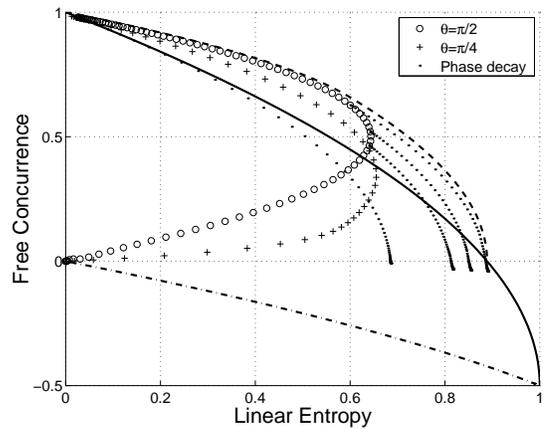}
  \caption{Evolution of ${\cal C}_{\rm free}(\rho )$ and
  $S_{\rm L}(\rho )$ for initial states
  (\ref{eq:rotation}) with $\theta =\{\pi /4,\pi /2\}$,
  $N=3.1$, and $\bar{n}=0$.
  The dotted curves show evolution produced by
  phase decay turned on after application of the unitary
  transformation $U$ to a selection of states from the
  $\theta=\pi/2$ curve. For this evolution, amplitude coupling is disabled,
  but $\eta\neq 0$ and $\bar{n}=1$.}
  \label{fig:superdecay}
\end{psfrags}\end{figure}

States above the Werner line can also be
generated by initially preparing the separable pure superposition state
\begin{eqnarray}
  \ket{\Psi (0)}=
  \brac{\cos(\theta/2)\openone_2+
  i\sin(\theta/2)\sigma_y}^{\otimes2} \ket{00} ,
  \label{eq:rotation}
\end{eqnarray}
with $0<\theta\le\pi/2$,
and then applying the effective reservoir
interaction with strong, ideal squeezing. In this case,
the time evolution follows paths as shown in
Fig.~\ref{fig:superdecay}.
There is a small region of the plane above the Werner line which cannot
be reached by this method. However, this region can be accessed by
switching off the squeezed reservoir interaction ($\beta_{r,s}=0$) and
employing phase decay ($\eta,\bar{n}\neq 0$) from initial states prepared on
the $\theta=\pi/2$ curve of Fig.~\ref{fig:superdecay} and to which the
local unitary transformation
$U=(1/2)\brac{\sigma_{x}+\sigma_{z}}\otimes\brac{\openone_2-i\sigma_{y}}$
(requiring single-atom addressing with
appropriate laser Raman pulses) has first been applied.
This is also illustrated in Fig.~\ref{fig:superdecay}.

We turn now to practical issues associated with our scheme.
Analysis of (\ref{eq:me}) shows that
the slowest rate featuring in the dynamics
is $(4\beta^2/\kappa )(2N-2|M|+1)$, which exhibits the
characteristic inhibited decay associated with atomic damping by a
squeezed reservoir \cite{Gar86}.
Meanwhile, inclusion of atomic spontaneous emission effects
(due to finite excited state populations) into the model
reveals characteristic rates $\gamma_j(\Omega_j^2/2\Delta_j^2)$
($j=r,s,t$).
Taking the rate for $j=r$ to be the maximum, and setting
$\bar{n}=0$, the
condition that spontaneous emission be negligible during the
state preparation period reduces to
$2g_r^2/(\gamma_r\kappa )\gg [1-\sqrt{N/(N+1)}\, ]^{-2}$.
This amounts to the condition of strong coupling cavity QED, made
somewhat more stringent however owing to the inhibited atomic decay
rate.
If we consider a recent cavity QED experiment for which the
parameters achieved were
$\brar{g,\kappa,\gamma}/2\pi= \brar{110,14.2,5.2}\textrm{MHz}$
\cite{Hoo00},
then for $N=2$ the above inequality reads as $332\gg 30$,
indicating that sufficiently strong coupling is experimentally realistic
for achieving significant levels of effective squeezing.
Furthermore, setting, e.g., $\Omega_r/\Delta_r=0.02$ and using
the above parameters, the characteristic state preparation time is
$\lesssim 50$~$\mu$s, which is orders of magnitude
less than single-atom trapping times in tightly-confining optical
dipole traps (see, e.g., \cite{Ye99,Sch00,Fre00,Sch01}).

\begin{figure}\begin{psfrags}
  \includegraphics[scale=0.4]{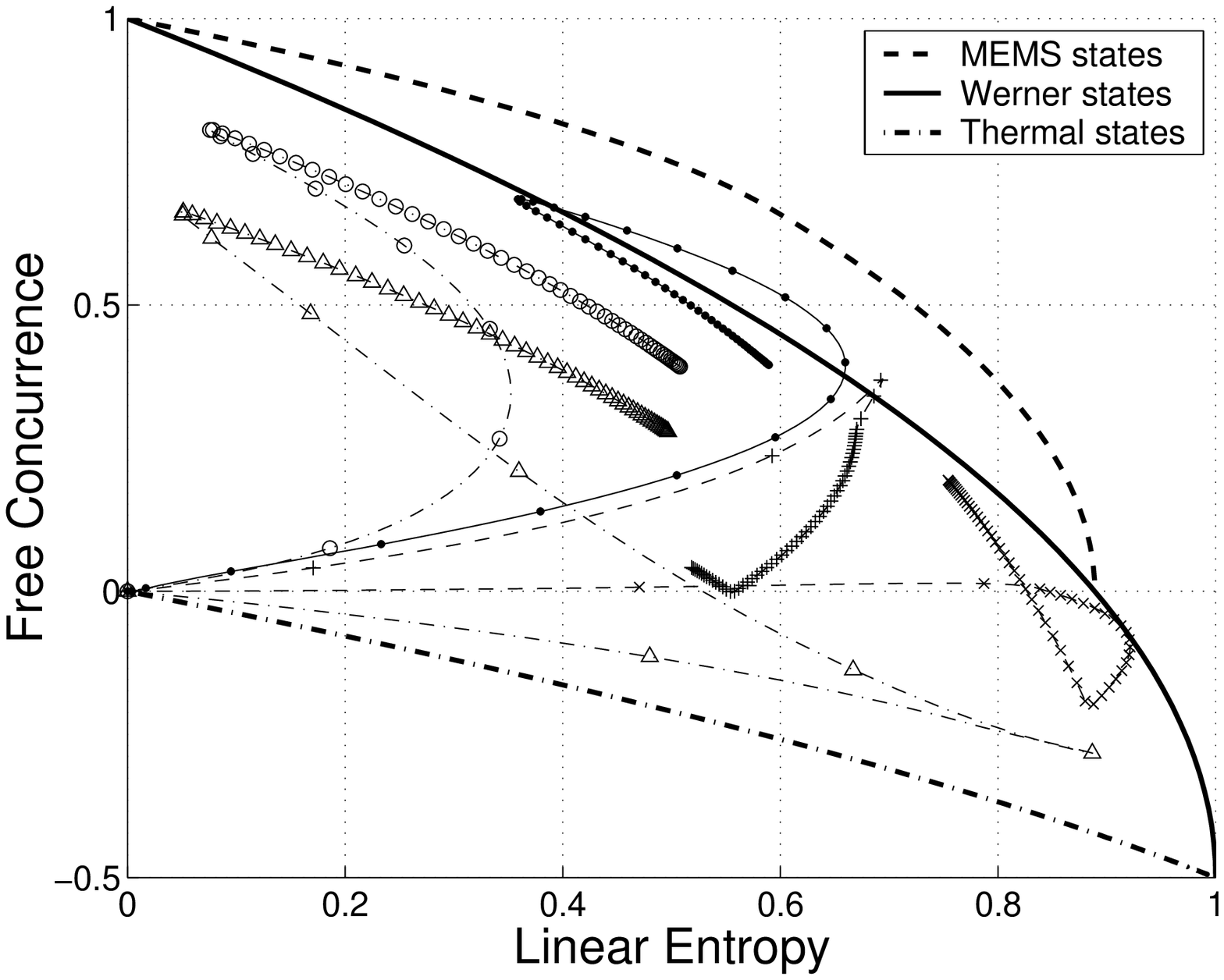}
  \caption{Evolution of ${\cal C}_{\rm free}(\rho )$ and
  $S_{\rm L}(\rho )$ with spontaneous emission effects for
  $\brar{g_r,g_s,\kappa,\gamma_j,\Omega_s,\Delta_j%
  }/2\pi=\brar{110,110,14.2,5.2,100,8000}\textrm{MHz}$ ($j=r,s,t$),
  $\Omega_t^2=\Omega_r^2-\Omega_s^2$,
  and $\bar{n}=0$, with
  the following initial states and values of $\Omega_r$:
  $\{$Eq.(\ref{eq:rotation}), $\theta =\pi /2$, $\Omega_r=120
  \; (N=2.3)\}$ ($\bullet$),
  $\{\ket{00},\Omega_r=173\; (N=0.5)\}$ ({\large $\circ$}),
  $\{\ket{11},\Omega_r=245\; (N=0.2)\}$ ($\triangle$),
  $\{\ket{01},\Omega_r=110\; (N=5)\}$ ($\times$),
  $\{\ket{01},\Omega_r=458\; (N=0.05)\}$ ($+$). Note that the points
  are not equally space in time.
  }
  \label{fig:sponemiss}
\end{psfrags}\end{figure}

In Fig.~\ref{fig:sponemiss} we present sample evolutions from
a model in which the excited atomic states have been adiabatically
eliminated, but in which effects of spontaneous emission plus
the cavity mode dynamics are included.
Using the cavity QED parameters quoted above, we see that a large
area of the linear entropy-concurrence plane can be
accessed. With the inclusion of spontaneous emission, decay
into the (weakly-coupled) state $\ket{\psi^-}$ can occur for
states with no initial projection onto $\ket{\psi^-}$.
This limits the maximal attainable concurrence and in
Fig.~\ref{fig:sponemiss} leads also, for the cases
with $\bra{\psi^-}\rho (0)\ket{\psi^-}=0$,
to a very slow decay
of ${\cal C}_{\rm free}(\rho )$ and increase in
$S_{\rm L}(\rho )$ after rapid initial
evolution to the optimal value of ${\cal C}_{\rm free}(\rho )$.
Note again though that the evolution can be frozen at any
point by simply turning off all of the light fields.
Note also that for the highest value of ${\cal C}_{\rm free}(\rho )$
attained in Fig.~\ref{fig:sponemiss}, one finds
$\bra{\phi^+}\rho\ket{\phi^+}\simeq 0.9$.
Higher cavity finesses (i.e., lower $\kappa$ values) improve the
performance of the scheme further and should be
accessible experimentally \cite{hjk}.

An alternative scheme for engineering the entropy and concurrence
can also be formulated using actual squeezed light to drive the
cavity and just a single cavity-mediated Raman transition.
Further, with nondegenerate squeezed light fields driving
distinct cavities it is possible to engineer the
states of distantly separated atoms \cite{Clark02}.

The scheme presented here can also be used to
prepare spin-squeezed
states of a larger number of cavity-confined atoms \cite{Aga89}.
While preparing this manuscript, we became aware of proposals
for preparing such states which use essentially the same
configuration as in this work, only not in the
guise of quantum reservoir engineering \cite{And01}.

In conclusion, we have proposed a scheme for engineering atomic
two-qubit states with any allowed combination of linear entropy and
concurrence. This should open the door to detailed experimental
investigation of purity and entanglement in quantum
information protocols.

\begin{acknowledgments}
We acknowledge helpful discussions with S. Rebi\'{c} and W. J. Munro.
This work was supported in part by the Marsden Fund of the Royal
Society of New Zealand.
\end{acknowledgments}


\begin{thebibliography}{13}

\bibitem{Whi01}
A.~G. White, D.~F.~V. James, W.~J. Munro, and P.~G. Kwiat,
Phys. Rev. A {\bf65}, 012301 (2001).

\bibitem{Mun01}
W.~J. Munro, D.~F.~V. James, A.~G. White, and P.~G. Kwiat,
Phys. Rev. A {\bf64}, 030302 (2001).

\bibitem{Lut98}
N. L\"{u}tkenhaus, J.~I. Cirac, and P. Zoller, Phys. Rev. A {\bf57},
548 (1998).

\bibitem{Pal89}
G.~M. Palma and P.~L. Knight, Phys. Rev. A {\bf39}, 1962 (1989).

\bibitem{Aga89}
G.~S. Agarwal and R.~R. Puri, Opt. Commun. {\bf69}, 267 (1989);
Phys. Rev. A {\bf41}, 3782 (1990).

\bibitem{Wer89}
R.~F. Werner, Phys. Rev. A {\bf40}, 4277 (1989).

\bibitem{Hag97}
E. Hagley {\it et al}., Phys. Rev. Lett. {\bf79}, 1 (1997).

\bibitem{Zhe00}
S. Osnaghi {\it et al}., Phys. Rev. Lett. {\bf87}, 037902 (2001).

\bibitem{Ple99}
M.~B. Plenio, S.~F. Huelga, A. Beige, and P.~L. Knight,
Phys. Rev. A {\bf59}, 2468 (1999).

\bibitem{EoF}
Note that the concurrence ${\cal C}$ is monotically related to
the entanglement of formation $E_{\rm F}(\rho )$, i.e.,
$E_{\rm F}(\rho )=h((1+\sqrt{1-{\cal C}^2})/2)$ where
$h(x)=-x\log_2x-(1-x)\log_2(1-x)$.

\bibitem{Gar86}
C.~W. Gardiner, Phys. Rev. Lett. {\bf56}, 1917 (1986).

\bibitem{Hoo00}
C.~J. Hood {\it et al}., Science {\bf287}, 1447 (2000).

\bibitem{Ye99}
J. Ye, D. Vernooy, and H.~J. Kimble, Phys. Rev. Lett. {\bf83}, 4987
(1999).

\bibitem{Sch00}
R. Scheunemann, F.~S. Cataliotti, T.~W. H\"ansch, and M. Weitz,
Phys. Rev. A {\bf62}, 051801(R) (2000).

\bibitem{Fre00}
D. Frese {\it et al}., Phys. Rev. Lett. {\bf85}, 3777 (2000).

\bibitem{Sch01}
N. Schlosser, G. Reymond, I. Protsenko, and P. Grangier,
Nature {\bf411}, 1024 (2001).

\bibitem{hjk}
H.~J. Kimble, private communication.

\bibitem{Clark02}
S.~G. Clark and A.~S. Parkins, unpublished.

\bibitem{And01}
A. Andr\'e, L.-M. Duan, and M.~D. Lukin, quant-ph/0107075;
A.~S. S{\o}rensen and K. M{\o}lmer, quant-ph/0202073.

\end{thebibliography}
\end{document}